\documentclass[10pt]{article}
\newcommand{\be}{\begin{equation}}
\newcommand{\ee}{\end{equation}}
\newcommand{\rf}[1]{(\ref{eq:#1})}

\begin{document}
\title{
The   Poincar\'e  Conjecture  and \\ the Cosmological Constant}
\author{M. D. Maia\thanks{Maia@unb.br}\\Instituto de F\'isica, Universidade de Bras\'{\i}lia,  DF 70910-900, Brazil.
}

\maketitle
\begin{abstract}
    The  concept of   deformation of  Riemannian geometry  is  reviewed,  with
 applications  to   gravitation  and  cosmology. Starting  with an  analysis of the
cosmological  constant problem, it  is   shown  that space-times  are deformable  in the
sense  of local change of  shape. These  deformations  leave    an  observable  signature
in the space-time,   characterized  by  a  conserved tensor,  associated  with  a  tangent
 acceleration,   defined by  the  extrinsic  curvature of the  space-time.    In  the
 applications  to   cosmology,  we  find  that  the   accelerated  expansion of  the
 universe   is  the    observable effect of   the deformation,  dispensing  with the
 cosmological  constant  and its  problems.


\end{abstract}


\section{The  Cosmological  Constant  Problem}

The  cosmological  constant   $\Lambda$  was introduced and  removed   by
Einstein in  1917, basically  because its   presence  is not  compatible with
the  Minkowski   space-time  as  a solution  of his  gravitational  equations. Yet,
the cosmological  constant  remains  as  the  basic  explanation  for the  observed
acceleration of  the universe,  within the  so called  $\Lambda$CDM  paradigm.

The usual  physical  interpretation  of    $\Lambda$  is  that  of the   vacuum energy  density $<\rho_v>$ of  quantum  fields. This follows from the   semiclassical  equations  with
  $\Lambda$ \cite{Zeldowich}:
\[
R_{\mu\nu}- \frac{1}{2}R g_{\mu\nu}  -\Lambda g_{\mu\nu} =-8\pi G  <\rho_v> g_{\mu\nu}
\]
These  equations  holds   true only under the  condition that
  $\Lambda g_{\mu\nu}$ in the left hand side cancels  with the $8\pi G <\rho_v> g_{\mu\nu}$   in the  right hand  side.
In  this case   we  obtain an  equation  compatible  with  the  Minkowski tangent solution, with  the  Poincar\'e symmetry and the  usual  quantum  field theory  (QFT)  required for  the evaluation of   the  vacuum  energy.

As it  happens,  the  theoretical estimates   give  the value
$$<\rho_v>  \approx 10^{76}  GeV^2/c^4$$
On the  other hand,  the acceleration of  the universe  measured by  the
 CMBR  in   various  precision  experiments,    indicates   that
$$ \Lambda/8\pi G \approx  10^{-47} GeV^2/c^4 $$
and  there is  no known procedure  in  QFT  capable of adjusting  these
values \cite{Weinberg,Weinberg:1}:

 Since the   cosmological  constant  term is   the  only  allowed addition
 to Einstein's  tensor,  following the   contracted  Bianchi  identity,  regardless
of  the existence   and the nature  of  the energy-momentum tensor, we  conclude that
the  cosmological  constant term in Einstein's  equations  is a  quantity  of  geometric nature
 which is   independent  of the right  hand  side  of  Einstein's  equations.

That  conclusion implies that  the  difference  between    $\Lambda$   and  $<\rho_v>$
is not  only  numerical,  but     mainly   conceptual:  \emph{The presence of  $\Lambda$
 does not depend on  the  nature of  what is placed  in the right hand  side of Einstein's
 equations. } Such  conceptual  difference  is a  consequence of  the   topological
 difference    between  the  Minkowski  and  the de Sitter  space-times  with their
 respective  groups of  isometries. As  we  know, it is possible  to  make  a group
  contraction  from the latter  to the former, but  at  the  cost  of the limit
$\Lambda \rightarrow 0$. However,   this limit is  not followed  by the  continuous
change of  shape of the respective  space-times. As  a   sphere  cannot be   be continuously
transformed into  a  plane   without  stretching  and  tearing off  the  manifold, the de Sitter
group  contraction  is not followed  by  a  smooth   deformation of  the  respective  geometries.
The  recent  solution  of  the  Poincar\'e  conjecture  on the  deformation of Riemannian
 manifolds  and  its   generalization   provided by Nash's  theorem, offers a  possible to  topological  solution of  the  cosmological  constant problem.

\section{The  Poincar\'e  Conjecture}

In  1904   Henry  Poincar\'e    conjectured  that  a  3-dimensional
(compact   and  simply  connected)    manifold  can    be     continuously  deformed
into  a   3-sphere. Although  it is  a very  intuitive   problem   it   was
 formally  proved  only recently  in 2006  by Grigori Perelman,   using  the
 Ricci flow  equation derived  by  Richard Hamilton in
 1982 \cite{RHamilton,Perelman:2},  given  by
\begin{equation}
R_{\mu\nu}= -\frac{1}{2}\frac{\partial g_{\mu\nu}}{\partial y}  \label{eq:Ricciflow}
\end{equation}One  simple  way  to derive   this  equation is  to write the   Ricci  tensor as
\[
R_{\mu\nu}=  (log\sqrt{g})_{,\mu\nu} -\Gamma^\rho_{\mu\nu,\rho}  +
\Gamma^\sigma_{\mu\rho}\Gamma^\rho_{\nu\sigma}
-\Gamma^\rho_{\mu\nu}(log\sqrt{g})_{,\rho}
\]
From  which,  using    geodesic  coordinates,  we  obtain the  the  Ricci   scalar
\begin{equation}
R=  g^{\mu\nu}R_{\mu\nu} =  \nabla^2 (log\sqrt{g})  \label{eq:Riccigeodesic}
\end{equation}
On the other hand,   replacing     $u=log \sqrt{g}$  in the Fourier heat  equation
$\nabla^2  u  =\frac{\partial  u}{\partial  t}$  we  find  that
\begin{equation}
\nabla^2 ( log\sqrt{g})=  g^{\mu\nu}\frac{1}{2}\frac{\partial g_{\mu\nu}}{\partial t}
\label{eq:aux1}
\end{equation}
Comparing the  right hand side of   \rf{Riccigeodesic} with the  left hand  side of
\rf{aux1},   removing the  trace   and     replacing   $t$  by   an  arbitrary  coordinate
$y$,     we obtain   the Ricci flow condition   \rf{Ricciflow},  up  to     a  sign and to
an arbitrary anti-symmetric tensor.

As  it happens  with  Einstein's  equations,  the Ricci flow  is  not  native  to
Riemannian  geometry,  but  it   represents   an  additional   postulate describing
 the  variation of  the  metric along a  given  direction,  not necessarily  time-like.

The  merit of  the  Hamilton/Perelman's   results  lies  in the interpretation of Fourier's
 heat  flow  in terms of  the continuous  deformation  of   the geometry  which  can be
exemplified  as  follows:
Consider the example of  a closed surface  $S$   being orthogonally crossed  by  a  number  of  flow  lines
per unit of  area, originating from the   heat  flux  in  a  heating  body.  Then, draw
another   surface,  for  example  a spherical  surface, $S_0$  inside  $S$,  such  that
 the  number of  flux   lines   per  unit of  area orthogonally crossing  it,  is  the
same  as  that for $S$.
Next,    freeze   these   flux lines, and  use   them  as  guides   to  smoothly  deform
$S$, always keeping it orthogonal  to  the  flux  lines,   without breaking  the  surface
or  even  without making  wrinkles,  until  reaching  $S_0$.  Under  such  conditions  we
  say   that  the   surface  $S$  has been  smoothly  deformed  into    $S_0$,  in accordance
with  \rf{Ricciflow}.

It is  a  simple matter   to see that the  Ricci  flow   is  not  compatible   with
 general  relativity: Writing Einstein's  equations as
$$
R_{\mu\nu}=  8\pi G ( T_{\mu\nu}-\frac{1}{2}Tg_{\mu\nu})
$$
and  comparing    with   \rf{Ricciflow},  we   obtain
$$
\frac{\partial g_{\mu\nu}}{\partial  y}=
-16\pi G  (T_{\mu\nu}- \frac{1}{2}T g_{\mu\nu})
$$
with the inevitable  conclusion that   the    Ricci  flow forces the
Einstein gravitation to propagate   linearly on any  direction $y$
of  space-time,  which
does  not  make sense in general relativity, even  considering only its  linear
approximation.  Either  we  use the hyperbolic  Einstein's  equations  or the
Parabolic Ricci flow equation  \rf{Ricciflow}.
 The  so called  Ricci-flow  cosmology   would be   an  entirely  new  proposition \cite{Carfora},

In the following  we   briefly  review  a  more  general and  older  concept of
deformation in Riemannian  geometry which is  compatible  with   general  relativity
and with the   present   cosmology,    where instead of   the  surface  $S$   we  have
a solution of  Einstein's  equations.

\section{The   Nash Geometric  Flow}

Perelman's   demonstration of the  Poincar\'e conjecture  suggests  the necessity of   a
mechanism  capable of  modifying  the  shape  of  a  manifold, that could   be  universally
applied to  all  Riemannian geometries, without imposing   a  constraint  to  the  metric,
independently  of   its dimension  and   metric  signature,
Such    general deformation process  was   derived by  John Nash  in  1956 and it is  given by
\begin{equation}
k_{\mu\nu}  = - \frac{1}{2}\frac{\partial   g_{\mu\nu}}{\partial  y}  \label{eq:Nashflow}
\end{equation}
where  $k_{\mu\nu}$  denotes the  extrinsic  curvature of  the   embedded Riemannian geometry.

The   expression  \rf{Nashflow} was proposed by  James  York,  restricted  to  3-dimensional
space-like surfaces evolving in a  space-time,
as  a   way  to  establish  the initial  value  conditions  in   the    ADM    formulation
of   general  relativity. As  we know,  the ADM proposition
did  not  work  for its  intended purposes, namely   to define  a canonical  formulation of
general  relativity  and  its  eventual use   to  quantize  gravitation, because    its
 incompatibility   with the  diffeomorphism  invariance of   general  relativity,  which
leads  to a constrained  Hamiltonian.  Even   using    Dirac's  procedure  for
constrained  systems  it  did  not  work because  the  Poisson bracket  structure is  not
covariant   with the  group  of  diffeomorphisms.

The     diffeomorphism  invariance  of  general  relativity is  a  statement about  which
observers are   allowed  in the  theory,  namely all.  As such,  the  diffeomorphism  invariance
has  to  do  with  the observable-observer   relation  and therefore it is  a  four-dimensional
characteristic. This means   that    in an  embedded  space-time  the  diffeomorphism
invariance must  remain  confined to the  four-dimensional  space-time  where  the observers and
their  gauge  fields  are  defined. In other  words, there is  no physical  reason  to assume
that it  would  extend  to  the  extra  dimensions. Hence, in an  embedded  space-time
the  diffeomorphism  invariance  holds in  the  four-dimensional  subspace  only.

For  some  authors,  the    embedding of the  space-time  into another  manifold is considered
to be  just as a  mathematical property,  without  physical  significance \cite{MTW}.  This
is  not  true  as  we  shall see.  Indeed  the   embedding  of  Riemannian   manifolds has its
origins  in  the solution of local  shape  problem.   That is,   in the  inability   of the
Riemann  curvature tensor  to  specify  the local  shape  of the  manifold \cite{Schlaefli}.
In the following, we  will show  that  an  isometric   embedding  of the  space-time
 produces  an  observable  effect.  In particular  for a  space-time  it can be  detected  as
a  physical   observable.

The  original  derivation of   \rf{Nashflow},   given in part  A  of  Nash's   paper was
unduly  complicated  by the  use of   smoothing operators \cite{Nash}.   To  show it  in  a
simpler  form,  we may  start  with the   case of  only  one  extra  dimension. The case of
 many extra dimensions  was  considered  in \cite{QBW}  leading  to  the  same  conclusion.

The  concepts of  flux  and  flow  lines introduced  by  Fourier  can be   formally  defined
as   the  trajectories  of  a  one-parameter  groups of  diffeomorphisms on  a Riemannian
sub manifold  $S$,  embedded   into  another   $V_D$,   defined as  follows:
Given   a point $p  \in  S $  and  the unit normal  vector $\eta$  at  $p$,  we  obtain  a
curve (called the   orbit  of  $p$)  with  parameter  $y$,
$\alpha (y) = h_y: S  \rightarrow V_D$,  orthogonal  to  $S$,  with velocity  vector
$\alpha' (y) =\eta$.
The   diffeomorphism   property  means  that    these orbits   can be generated by
infinitesimal   increments   of  the  parameter  $y$,  so  that  subsequent points  are
defined by  a  composition law   given by $ h_y (p)\; o \; h_y' (p) =h_{y+y'}(p)$.
The group  property  follows from the  definitions $ h^{-1}(p)=h_{-y}(p)$  and  $ h_{0}(p) =p$.
 Given a  geometrical  object $\Omega$  defined in  $S$,  it  propagates along  these orbits
 by the Lie  transport \cite{Crampin}
$$ \Omega' = {\Omega}   +  y \mbox{\pounds}_\eta \Omega$$

\section{Embedded  Space-Times}

Denoting  the  metric  of the  five-dimensional embedding  space $V_5$ by
$\mathcal{G}_{AB},\;\;  A, B  =1..5$,   the  isometric  embedding of  a
Riemannian   geometry  $S$    a   map  $X:S\rightarrow V_5$,  such that
\begin{equation}X^A{}_{,\mu} X^B{}_{,\nu}{\cal G}_{AB}= g_{\mu\nu}, \;
X^A{}_{,\mu}\eta^B {\cal G}_{AB}=0, \;
\eta^A \eta^B {\cal G}_{AB}=1 \label{eq:X}
\end{equation}The  four-dimensionality of  space-time  is  a  consequence  of
the dual   properties  of  gauge  fields,  which  consists of our   main observational
tools. By  definition,  the  embedding of  a space-time     maintains  that  dimensionality,
but it adds a  topological  information by the  extrinsic  curvature. This  is   a
measure  of the  variation of  the normal  vector $\eta$ when its  foot  is  displaced
in a  tangent  direction to $S$, projected into $S$:
\begin{equation}\eta^A_{,\mu}  =  -{k}_{\mu}{}^\rho   X^A_{,\rho}
\end{equation}
To understand  the meaning of  this  extrinsic  curvature, we may   picture  the orbits
of   points  in  $S$   with tangent  vector $\eta$  as plane curves in $V_5$,    whose
acceleration  $\eta'$ is   orthogonal to  $\eta$ (following  the Frenet  equation).
Therefore, it follows  from \rf{X}  that  $\eta'$   is  a  vector  tangent  to  $S$. From
the above expression  it  follows  that   the extrinsic  curvature  represents an
 acceleration  in  $S$,    giving  a   mechanical   meaning  to  Nash's   deformation.

The  vectors $\{X_{,\mu}\}$ define  a  tangent basis  to $S$, so  that
$\{X^A_\mu ,  {\eta}^A_a  \}$   define  a Gaussian  frame of  $V_5$.
To obtain  \rf{Nashflow}  consider  the    Lie  transport of  that   Gaussian frame
along the orbits,   giving    a new  set  of  vector fields
\begin{equation}
X'^A\!\!  =  X^A +   y \mbox{\pounds}_{\bar{\eta}}X^A
=  X^A{}_{,\mu} +  y \;  \;\; \eta^A{}_{,\mu},
\;\; \eta'^A  =\eta^A  +    y\;\mbox{\pounds}_\eta \eta =\eta^A \label{eq:X'}
\end{equation}
Nash's  local  embedding theorem   consists in  showing that
$X'^A$      define   a new \emph{deformed  manifold}  $S'$ provided   they
satisfy  similar isometric  embedding   equations
\[
X'^A{}_{,\mu} X'^B{}_{,\nu}{\cal G}_{AB}=g'_{\mu\nu},\; \; X^A{}_{,\mu}\eta^B {\cal G}_{AB}=
0,  \;\;  \eta^A \eta^B {\cal G}_{AB}=1
\]
where   $g'_{\mu\nu}$  is  the  deformed  metric.
Replacing $X'^A{}_{,\mu}$   given by  \rf{X'} in these  equations we obtain in a
straightforward  way that
\begin{eqnarray*}
&&g'_{\mu\nu} =   {g}_{\mu\nu}-2y{k}_{\mu\nu} + y^2 {g}^{\rho\sigma}{k}_{\mu\rho}{k}_{\nu\sigma}\\
&&k'_{\mu\nu}  ={k}_{\mu\nu}  -2y {g}^{\rho\sigma} {k}_{\mu\rho}{k}_{\nu\sigma}
\end{eqnarray*}
Deriving   the  first  equation  with  respect  to  $y$,  and  comparing  with   the second
equation    we obtain Nash's  geometric  flow  \rf{Nashflow}.
Notice that in  each  deformation obtained  by \rf{Nashflow},  the extrinsic  curvature
is  independent of  the  metric,   satisfying the  Gauss-Codazzi  equations  for  the
embedded   manifold. Thus,  for any particular  deformation the   variable  $y$  is
contained  in the  expression of  the extrinsic  curvature.

The isometric  condition   \rf{X} means  that  the  geometry  of  the  embedded  space
is  induced  by  the  geometry  of  the  embedding space
(so  that  we  have  only one  metric  geometry).  Therefore, to  obtain the  same
gravitational  field  definition,  the
metric of  the    embedding  space must be  defined by the  same Einstein-Hilbert
variational principle,  leading  to  the
  the  higher  dimensional   Einstein's  equations (in arbitrary coordinates)
\begin{equation}
^5{\cal R}_{AB}  -\frac{1}{2} \,  {^5{\cal R}} {\cal G}_{AB} =G_* T_{AB},
\;\;\;  A,B ... 1..5  \label{eq:BE0}
\end{equation}
where $G_*$ is  the   appropriate   gravitational  constant   (including  the  solid
 angle) for  a  5-dimensional  space \cite{ADD}.

To  obtain the  four-dimensional   equations  we  only  have  to  write  the  above
equations  in the  Gaussian  frame  $\{X^A_{,\mu}, {\eta}^A \}$  of   the embedding
space.\footnote{Notice  that  this  projection  can  be  made in any   four-dimensional
space-times  regardless of  specifying the  value of  $y$.}  We  obtain  two sets
of  equations:
\begin{eqnarray}
&&R_{\mu\nu}-\frac{1}{2}R g_{\mu\nu}   - Q_{\mu\nu}=
-8\pi G T_{\mu\nu}\; \hspace{2mm}  \label{eq:BE1}\\
&&k_{\mu;\rho}^{\;\rho}-h_{,\mu} =0,\;\;\;\;\;  \mu , \nu  =1..4 \label{eq:BE2}
\end{eqnarray}
where  we  have  denoted the  tensor  quantity
\begin{equation}Q_{\mu\nu}  =  k_\mu^\rho k_{\rho\nu} -hk_{\mu\nu}  -
\frac{1}{2}(K^2-h^2)g_{\mu\nu}  \label{eq:Qmunu}
\end{equation}It  follows directly  from this  definition  that
\begin{equation} Q^{\mu\nu}{}_{;\nu}=0
\end{equation}so that  in principle   $Q_{\mu\nu}$  corresponds  to  an
observable  in  the space-time, in the sense of  Noether's theorem.

It  should  also be  noted  that the  definition of  the   embedding  geometry  by the
higher  dimensional  equations   \rf{BE0}  provides  a more  general embedding  than
the common practice of    specifying  a particular  5-dimensional     embedding    space,
such  as  a  flat  space,   or  a  de Sitter/anti-de Sitter   spaces,  or  a  Ricci-flat
embedding  space. Since the  metric  of the  embedded  space-time  is  induced by that
of the embedding space, these  particular  choices,  imply in  a  constrained  embedding.
 For  example,  when  the  5-dimensional  embedding  space is  flat,  the Gauss-Codazzi
embedding  equations  do not  have  a  solution for  specific  metrics \cite{Kasner,Szekeres}.
In the case of  Ricci-flat   5-dimensional  embedding  space, we
obtain   from  \rf{BE0} that  the Ricci tensor of the  space-time is not  zero but it
must be  conserved \cite{Campbell,Romero}. Finally in  the Randall-Sundrum  brane-world
 models  the extrinsic  curvature  is   not dynamical, but rather   given  by  the
 four-dimensional confined sources.

  On the  other hand,  in the   geometric  deformation of  Nash  with  the
Einstein-Hilbert  principle,  the  four-dimensional  gravitational equations  is  defined
by    the  same   four-dimensional  Einstein's  tensor  plus the   additional   tensor
$Q_{\mu\nu}$  in place  of  the cosmological  constant.

Since   $Q_{\mu\nu}$ is  conserved,  it must  correspond  to an observable  quantity.
Indeed,    the  standard Friedmann-Lemaitre-Robertson-Walker (FLRW)  universe is  described
by the classical   Friedman's equation
\[
\dot{a}^2 +  k \approx   -\frac{8 \pi G}{3}\rho a^2
\]
However,  for  the  FLRW  universe   embedded in a  five-dimensional  space  defined
by  \rf{BE0},  Friedman's   equations becomes
\[
\dot{a}^2 + k  \approx
-\frac{8 \pi G}{3} \rho a^2 + \frac{b^2}{a^2}
\]
where we  have  denoted   $a(t) = g_{11}$  and  $b(t)  = k_{11}$.  The  term  $b^2/a^2$
corresponds  to  the  deformation  tensor $Q_{\mu\nu}$.

In a  previous  publication   we  have  compared  the   effect of
$b^2/a^2$  with the  effect of  a hypothetic   phenomenological  fluid, the  x-fluid
idealized  by  Turner  and White  to  simulate  the gravitational effect of dark
energy \cite{GDE1}.  In a  subsequent publication  we  have  generalized  that   result
in two  ways: First, instead of  a  phenomenological  fluid  the  extrinsic  curvature
was  regarded  as  an  independent  spin-2  field,
second,  by  using a  model independent    statistical analysis
based on the luminosity-distance measurements.  In both cases  we    have  found  that
 the   deformation term  $b^2 /a^2$   is   sufficient  to  describe  the  acceleration
 of the universe.  More importantly,  we  have  found  that the  cosmological  constant
does not play any  significant  role  on this  deformation \cite{GDE2}.

One interesting  question  concerns  the  end  of  the  acceleration.
When   the  extrinsic  curvature   is  proportional  to  the  metric  we  find  a peculiar
situation:
Admitting that $k_{\mu\nu}  =\alpha_0 g_{\mu\nu}$,   the  space-time  contains  only  umbilical
points and it behave  as  a  constant  curvature  space.  In  such  cases   the   Nash
 deformation  ceases   and  \rf{Qmunu}  gives
\begin{equation}Q_{\mu\nu} =  3\alpha_0^2   g_{\mu\nu}
\end{equation}
Replacing these  conditions  in   \rf{BE2}  it  becomes  an identity  while  in
\rf{BE1}  we obtain  Einstein's  equations  with   a  cosmological  constant term,
representing the    end  limit of the
 deformation process.

\vspace{1cm}

We   conclude    that  the  inclusion of  a  cosmological
constant has   led  us   to    a  sequence  of  conceptual  and  observational  difficulties.
In  particular  its  comparison  with the  vacuum  energy  density  in  quantum field
theory cannot be  sustained, mainly  because  the  cosmological  constant  would  exist
independently   of  the   existence and  nature of  the  energy-momentum  tensor.
 By  comparing   such problem  with  the Poincar\'e   conjecture  we   conclude  that
 the  problem  could be  solved  if  we  include  in  Einstein's gravitational  theory
a smooth   deformation  mechanism  capable  of  modifying  space-time  topology.
Such  mechanism  exists  and  it is given  by  the  geometric   deformation  introduced  by
Nash in  1956,
based on the   foundations of  Riemannian  geometry.  It is  entirely compatible  with
 general  relativity, but   it adds  a  new  observable quantity  constructed  with  the
extrinsic  curvature  of  the  space-time.
When  this   is  implemented in the  FLRW  cosmology,  we  find  that
this  conserved  quantity  corresponds  to  the  observed
acceleration  of  the universe.  We   also  find  that  such   acceleration   ends  in a
constant  curvature  de Sitter  phase.

\end{document}